\documentclass[twoside, twocolumn]{article}

\usepackage[top=1.0in, bottom=1.3in, left=1.3in, right=1.3in]{geometry} 

\usepackage{ifpdf}
\ifpdf
\usepackage{thumbpdf}
\RequirePackage[colorlinks,hyperindex,plainpages=false,pdfauthor={Robert
  Primmer},citecolor=black,urlcolor=blue,linkcolor=black]{hyperref}
\else
\RequirePackage[plainpages=true]{hyperref}
\usepackage{color}
\fi

\usepackage{wrapfig}

\usepackage{subfig}

\usepackage{epsfig}

\usepackage{array}

\usepackage{longtable}

\usepackage{verbatim}

\usepackage{fancyhdr}

\newcommand{\ados}{distributed object store }
\newcommand{\adoss}{distributed object stores }

\newcommand{\aos}{object store }
\newcommand{\aoss}{object stores }
\newcommand{\apop}{principles of operation }

\makeatletter
\newcommand{\setversion}[1]{\def\@version{#1}}
\newcommand{\version}[0]{\@version}
\makeatother

\begin{document}

\setversion{2.1}
\title{Distributed Object Store Principles of Operation \\ The Case for
  Intelligent Storage}
\author{Robert Primmer}
\date{May 10, 2010}

\maketitle

\ifpdf
\pdfbookmark[1]{Abstract}{abstract}
\fi

\begin{abstract}

In this paper we look at the growth of \adoss (DOS) and examine the underlying mechanisms that guide their use and development. Our focus is on the fundamental \apop that define this class of system, how it has evolved, and where it is heading as new markets expand beyond the use originally presented. We conclude by speculating about how \aoss as a class must evolve to meet the more demanding requirements of future applications.

\end{abstract}

\section{Introduction} \label{s:intro}

While the concept of \aoss has existed in the academic realm for over a decade \cite{Venti, Val}, there have been relatively few commercial implementations.  In the first decade of the 21st century a handful of commercial offerings emerged with the enterprise sector as their initial target market. As we enter the second decade we see use of \aoss generalized and expanded to suit the Cloud use case.

To successfully move from concept to commercial product requires that you solve a business problem. For new products success is typically achieved either by creating a whole new market or by taking an existing problem and trying to do it better (a/k/a optimization), with the latter being the approach taken more frequently as it's usually easier to improve upon existing work than create something completely new from whole cloth.

For commercial \aoss the initial target of opportunity was to provide a spinning disk alternative for business data archived to slower media (such as tape or optical).  The primary value proposition was conceptually simple: the globalization of business in the 21st century meant that the pace of business was increasing at an increasing rate \cite{economy}, thus the need for rapid and ready access to the data upon which business relies had to increase accordingly.

Whatever their other values, tape and optical were designed to efficiently write data \emph{sequentially}, thus these media were destined to fail the first test of rapid \emph{random} data retrieval.  They failed the second test of ready access due to the mechanical, and sometimes human, element in the data retrieval process itself.  Before the data can be read, the medium must first be located and loaded through some combination of human and/or mechanical process; thus it can take minutes to retrieve a single file.  While there are ways to improve time to first byte (TTFB) through caching and read pattern optimizations, these media simply weren't designed to be efficient at random data access and were therefore never as fast as spinning disk for random read patterns in the general case.

As tape and optical were most frequently used for data archival, the first commercial \aoss were built with data archive as their initial design center.  It's important to note that that there's nothing intrinsic to objects, and by extension object stores, that limits their application to data archive.  However, commercial \aoss came about at a time when a pressing business problem to be solved was to make archived data more useful to the business by substantially reducing the time required to randomly read stored data -- essentially reducing TTFB by 1 to 2 orders of magnitude.

This was not the only business problem proffered as the rationale for changing the medium of archive from tape/optical to spinning disk. The second most common rationale provided was data durability. The nature of archive data is that the storage time horizon changes from months to decades. Since tapes tend to degrade over time \cite{tape} and optical formats frequently change \cite{optical}, the second value a spinning disk solution brought was the notion of easy and continual upgrade to new and denser HDD/SDD. This not only solves the problem of medium obsolescence, which is itself a substantial problem if archive data is of any real value, but simultaneously allows customers to ride the attractive cost curve established with HDD, which has proven exceptionally beneficial to storage consumers\footnote{Many in technology are familiar with Moore's Law, which observes a doubling in processor speed roughly every 18 months. Less familiar is Kryder's Law, which observes a \textbf{50-million fold increase} in storage capacity since the introduction of the first disk drive by IBM in 1956 \cite{sciam1}. In fact, over the past four years, HDD areal density (measured as gigabits per square inch) has been doubling roughly every 11 months, whereas processor capacity has been doubling somewhat less than every 18 to 24 months \cite{sciam2}.}.

All these factors combined to form fertile ground for a new class of storage to emerge. And a market was born.

\subsection{Terminology} \label{s:terms}
\begin{description}
\item[DDB] Distributed Database
\item[DOS] Distributed Object Store
\item[HCP] Hitachi Content Platform
\item[HDD] Hard Disk Drive
\item[HDS] Hitachi Data Systems
\item[SSD] Solid State Drive
\item[TTFB] Time to First Byte
\end{description}

\subsection{Document Organization}

The remainder of the document is organized as follows. \S\ref{s:dospop} talks about the general \apop for \adoss as a category, examining them from a systems, market, and client view and describing the problems they seek to solve.  \S\ref{s:doschars} presents characteristics that are common to present-day DOS and discusses where they differ.  \S\ref{s:futures} looks at how object stores as a class must evolve to meet the more demanding requirements of future applications. \S\ref{s:summary} concludes with a summary of the topics covered in this paper.

\section{Distributed Object Store} \label{s:dospop}

\subsection{Operational Definitions}
In computer science the definition of terms are frequently overloaded, sometimes varying considerably. We begin this section by providing our operating definitions of some common terms used throughout this paper.

\subsubsection{Object} \label{s:objdef} The term object is purposefully generic so it can be applied to many things.  For our purposes we'll use the term \emph{object} to denote a data construct that has at least two constituent parts: data and metadata, where \emph{data} represents the client data and \emph{metadata} represents an arbitrary set of information that is in some way connected to the (client) data.  Therefore, an object minimally equals the union of data plus metadata.

Note that from the client's perspective, each discrete object is essentially atomic when viewed as a storage element. However, from the perspective of the object store, a single user object can result in many fragments, possibly dispersed throughout one or more clusters that constitute the full DOS.

\subsubsection{Distributed Object Store} \label{s:dosdef} 

An \emph{\aos} is a collection of loosely coupled objects that may or may not have relation to any other object residing within the same object store. At present there isn't a canonical structure for an object store such as one finds in a traditional hierarchical file system\footnote{An object-based storage device (OSD) specification exists and has been ratified \cite{t10} but has not seen widespread commercial use.}. However, some facade representing a structure recognizable by a human user may be presented, typically to allow end-user traversal of the object store.

The terms ``object store'' and ``distributed object store'' can essentially be used as synonyms as the distinction becomes one of distance. However, there is no universally accepted definition of how much distance must be maintained between clusters to be considered a ``distributed'' object store. At its simplest, if objects can be dispersed across of set of physically discrete hardware elements (such as nodes), the object store is distributed.  An additional qualification is sometimes applied where the distance between the hardware elements is expected to extend beyond the confines of a single data center, perhaps extending to different geographies.

For the purposes of this paper we'll use the simpler definition as that expands the set of solutions we can consider without requiring repeated qualification of the terms.

\subsubsection{Distributed Database} \label{s:ddb} 

Not surprisingly, there are several definitions for what constitutes a distributed or decentralized database. For our purposes we'll consider a database to be distributed if it follows the same principles described earlier for an object store. Perhaps the best known example of a DDB is DNS \cite{DNS}.

The concept of a DDB is important to a \ados because, at its core, a DOS is software built on top of a DDB. Something has to do the heavy lifting of keeping track of billions of discrete object fragments and coalescing these back into a cogent, atomic object usable by applications and humans alike, and that job falls primarily to the DDB. It is the strength of the DDB that will determine the strength of the DOS; therefore, the tolerance limits of the DDB will be the tolerance limits of the DOS itself. For example, the upper bound on the number of objects a DOS can handle is the number of objects the underlying DDB can handle.

In the case of hierarchical file systems, the burden of location awareness for each visible element that makes up the object collection (such as separate data and metadata files) is placed on the client. As there isn't an agreed file system construct for the notion of attaching arbitrary metadata to a user file (such as by using extended attributes \cite{EA}), it is incumbent upon the client to create multiple files to achieve this end, each of which must be individually accessed by means of a fully qualified pathname. With an object store, typically the burden of location awareness shifts from the client to the server \cite{RJP2}.

\subsection{Basic Model}

At its most basic any \aos can be viewed from the perspective of the client and from the perspective of the server. In this section we describe \aoss from these perspectives, but add a third perspective that deals with the market tensions that surround enterprise-class \aoss as these market conditions have significantly colored the present perception of \aoss as a class and have a direct impact on their future development.

\subsubsection{System View} \label{s:sysview}

From the system perspective the \aos looks similar to a file store: there are clients and servers, clients make data requests and servers service these requests. In the case of an object store the client is typically not a human user, but an application specifically written to interact with the DOS.  Some \aoss will front the core service by presenting different protocols to the client, such as CIFS and NFS. These front-end systems act as protocol converters, arbitrating the differing protocols used by clients with the protocol used by the server. For \aoss that operate in the cloud, such as Amazon S3 and Nirvanix, the protocol used to communicate with the \aos is typically HTTP and is often designed to qualify as a RESTful interface \cite{REST}.

The use of HTTP as the high-level communications protocol indicates one of the first differences between an \aos and a traditional storage subsystem, as TCP/IP is the protocol of choice, versus block-based protocols such as Fibre Channel.  All this leaves an \aos looking suspiciously similar to a file store.  Both are often used for ``unstructured data'', which at its simplest is another way of saying ``file'' instead of database. Both are typically accessed via TCP/IP.  So what are some of the differences?

For starters, a file store is most commonly a file system of some variety exported for use on a LAN. A file system is just just another type of database and hence induces structure and structure induces limitations, such as the number of files that can be stored in a particular directory or file system. It also enforces a grouping where none may naturally exist based on the file content itself. With an \aos there isn't the same notion of order and hierarchy.  Instead the \aos is viewed as a flat namespace in which objects of various types are mixed together in a manner opaque to the client.  If structured storage is the china cabinet where dishes are neatly stacked and ordered by type and circumference, unstructured data is the junk drawer in the kitchen where random items are thrown together with no particular sense of connectedness a priori.

Even in cases where an \aos presents the facade of a file system to the human user, the objects themselves are scattered about the cluster in a manner uncontrolled by the client. So while \aoss ultimately provide a data storage repository, they are better understood as a software system that collects user data and performs some set of actions against that data while holding it on some form of persistent storage, which today typically takes the form of HDD.

Because it is ultimately a software system, the \aos can perform an arbitrary set of functions against the data, both during ingest and post-ingest, that go beyond the traditional storage functions of creating local and remote replica copies, applying access permissions, performing deduplication, etc. For example, the \aos may perform transformations on image files to present different quality images to different classes of users, or it may perform sophisticated data classification based on a set of criteria specified in the object metadata or in response to external events, such as reaching a certain time boundary or access frequency.

\subsubsection{Market Tension}

The fact that an \aos is really just a software system running on a cluster of servers means that the system can theoretically perform an unbounded range of function, and, properly designed, do so at spectacular scale. It's this quality that creates a natural market tension when \aoss are introduced.

Since the \aos itself can be extended to perform many, if not all, of the functions performed by existing content management software, there no longer is a crisp demarcation between the domains of the application software and the storage system.  Once they've conquered the base functions requisite for reliable data storage, it's a natural evolution for \aoss to begin to ``move up the stack'' and perform more and more functions that were once the exclusive domain of the application vendor.  Likewise, application vendors have taken note and are actively seeking to add more of the functions of the object store to their own software, as nobody wants to see their own product commoditized.

This tension has had the effect of dampening the growth of commercial object stores in the enterprise market.  However, it can be reasonably argued that from the perspective of the customer, it's better to have common functions performed in a common way in a single place.  Having $n$ applications provide the same functionality, each in distinct ways, raises IT cost as the cost to train personnel and manage these systems must increase linearly at best.

How this dynamic plays out over the next few years will be interesting. The wildcard in all of this is Cloud. If enterprise-class \aoss give only a glimpse of how and precisely where data fragments are stored and reassembled, Cloud makes this positively opaque. As Cloud data storage is built upon, and an extension of, the principles of an object store as presently used in the enterprise market, it is likely that the market will see a greater blurring of the line between ``application'' and ``storage'' over time.

\subsubsection{Client View}
In the world of \aoss clients are typically applications instead of human users\footnote{Technically speaking, the client is always an application, even in the case of a user individually storing files on a NFS or CIFS share.  However, for our purposes we apply some rounding to focus on how systems are commonly viewed.}. Even when there are human users acting against some common LAN file protocol such as CIFS or NFS, the client of the \aos is typically the piece of software performing protocol translation that sits between the human user and the object store. Thus, to the client the \aos is a software service that sits on the other end of a TCP/IP connection and responds to requests much the same as any other software service.

The degree of opacity of the \aos varies by implementation type. In the case of content addressable storage (CAS), the client is often returned only a completely opaque handle when submitting an object and is given no information about the storage of the object \cite{Rhea}. In other implementations, such as the Hitachi Content Platform (HCP), the client is presented a familiar file system semantic as facade to the object store and is returned a file handle upon object ingest.

Of course, there's no intrinsic relationship between the presentation layer exposed to the client and the manner in which the data is ultimately distributed and stored, but using a familiar facade does provide a means for both the application and the human user to traverse their data in a commonly understood fashion. It also allows for user-created logical groupings as a means of data organization. It doesn't matter that under the covers HCP doesn't group the data in similar fashion, instead choosing the most efficient means of spreading the data across the cluster, because the real purpose of the presentation layer is to help the end user better navigate the system\footnote{The exception to the general case is when a system administrator wants to assign a collection of objects to a particular class of back-end storage subsystem. In such a case, the groupings presented matter.}.

Every design model selected results in a set of tradeoffs. In \S\ref{s:diffs} we walk through some of the more important differences between the various design models of object stores.

\subsection{Problems to Solve}
To be successful an \aos needs to solve several problems; some are basic, but some are quite hard. In this section we survey the issues common to \aoss in general. We begin by briefly enumerating the basic problems common to all DOS implementations and then provide an expanded look at two of the harder problems to solve: scalability and concurrency.

\subsubsection{Basic Problems}
The more common basic problems to be solved are briefly described below, with a more detailed description in \S\ref{s:doschars}.

\begin{description}
\item[Multiple Entry Points]\hfill \\ The system must allow for multiple independent applications simultaneously performing operations such as read and write.
\item[Global Namespace]\hfill \\ An \aos should present a global namespace (GNS) to the client.
\item[Time Horizon] \hfill \\ The time horizon for an \aos can be decades.  Therefore, the system must be designed to periodically check the veracity of the data stored as all media degrade over time.
\item[Access Protocol] \hfill \\ The access protocol should work equally well over a WAN (such as the Internet) as over a LAN. Therefore the access protocol cannot be chatty, as most network file system protocols tend to be.  Further, it should support mobile devices, such as smartphones, as clients.
\item[Unstructured Data] \hfill \\ The system must be designed to optimize for unstructured data as this will be the predominant type of data over the next decade. \cite{idc1}
\item[Hardware Agnostic] \hfill \\ Hardware changes frequently. This includes servers, storage subsystems, and even the storage medium itself, as seen with the introduction of solid state drives (SSD) which provide better random I/O properties but present different challenges for long-term use (such as wear patterns). Given the rule on Time Horizon, it's imperative that an \aos be designed to be hardware agnostic. Like the storage medium itself, all supporting hardware must be fungible.
\end{description}

\subsubsection{The Bookkeeping Problem (Scale)} \label{s:bkp} 

In \S\ref{s:ddb} we said that at its core a DOS is essentially a DDB with additional software layered on top to provide value-added features.  Here we expand on that thesis by asserting that the DDB design is the single most important aspect of the whole DOS architecture; if you fail at the DDB design, the system will quickly reach maximum scale -- not by the amount of capacity that can be physically added to the cluster, but with the number of objects the system can simultaneously keep track of and therefore allow to be ingested. The bookkeeping problem is the principal gate to overall system scale when dealing with an object store, in both the capacity and performance realms.

Scale is simply a hard problem to solve in computer science. The nirvana of ``infinite scalability'' looks great on a marketing data sheet, but has thus far proven elusive in actual implementation. Making the problem harder still, with \aoss the Time Horizon problem means that scaling out needs to be essentially seamless because forklift upgrades are counter to the promise of an ``active archive''.

A true enterprise-class DOS needs to scale to 10's of billions ($10^{10}$) of \emph{user} objects\footnote{Note that in several existing implementations, a DOS takes a single user object and converts it into $n$ cluster objects, where at the very least $n \ge 2$ for simple mirroring, but more likely ranges to $2 \le n \le 14$ if more storage-efficient means are used for data protection such as erasure encoding \cite{Guru}. At this scale the total cluster objects used to store 10 billion user objects can quickly reach 100 billion cluster objects ($10^{11}$).}. With Cloud this number has the potential to increase by several orders of magnitude, so it's easy to see how difficult it becomes for a loosely coupled, distributed system to simultaneously keep track of cluster objects in the 100's of trillions ($10^{14}$). Relational databases (RDBMS) don't do well with table entries that range into the trillions, and even if they did, locating the multiple elements that constitute a single user object can grow in time quadratically.

There's no easy way to solve this problem with RDBMS technology. For starters, RDBMS software is optimized for ACID consistency which makes it suboptimal for distributed databases \cite{gray}. Further, the cost to operate such a substantial RDBMS system would be prohibitively expensive, as it would require best-of-breed commercial RDBMS software coupled with very fast (and therefore very expensive) storage systems just to run the RDBMS alone, easily eclipsing the total cost of the object store itself.

With the HCP product it was determined that the best way to handle this problem was to break the database up and distribute it more or less equally among the nodes of the cluster, much the way objects are distributed.  This model has several advantages:
\begin{itemize}
\item The number of objects each \emph{node} in the cluster can store increases, thus the resulting number of objects the \emph{cluster} can store grows quite large;
\item The time to locate a particular object fragment is quicker as the lookup operation itself is likewise partitioned; and
\item The protection model for the database itself can follow essentially the same model used to protect user objects within the cluster, thus leading to greater protection for the DDB itself -- which is crucial to the proper function of the whole object store, as described in \S\ref{s:ddb}.
\end{itemize}

As discussed earlier, every design model necessarily brings with it a set of tradeoffs. In this case the greater resilience and operational efficiency of breaking up the database and distributing it among the constituent nodes of the cluster leads to the issue of concurrency.

\subsubsection{Concurrency} \label{s:concurrency} 

Like infinite scale, concurrency is another of the well known ``hard problems'' to be solved in computer science. The fact that we're talking about \emph{distributed} object stores means that the concurrency issue has to be solved in order to have a viable solution at large scale. Breaking up a database to manageable parts and distributing the pieces among the nodes of the cluster may solve the scale problem, but it's still a failure if the individual databases all have differing views of the truth.

With very large-scale systems dispersed over distant geographies, synchronous concurrency isn't practical. There are a number of models used to provide suitable concurrency. Amazons S3 uses the ``eventually consistent'' model, which allows geographically dispersed sites to be inconsistent, but only for a period of time considered sufficient for the solution provided \cite{aws}. How long it's acceptable for the same object to be in different states is, of course, a function of the needs of the client application.

While there are numerous dimensions to the problem, we can generally state that there's an inverse relationship between the economics of the solution and the time it takes to reach perfect synchronicity.  Applications that require very fast synchronization, such as those seen in the financial sector with bank transactions and stock market trades, require the customer to pay a premium for systems that reach synchronicity very quickly. However, for a large swath of applications such cost is unnecessary to adequately meet the client application needs.

In their initial instantiation, ``Cloud'' applications tend toward this latter set of consumers in part because the Internet is the medium of choice for data transport. Since the Internet is primarily constructed as a packet-switched network where best effort is the the accepted modus operandi, using it as the transport medium acts as a limiter. With the Internet there are no hard guarantees of consistent packet speed or ordering end to end. And while there are research ideas on how to improve on known problems such as ``middle mile'' congestion \cite{Akamai}, the state of the Internet today is a network built for ready global access at affordable cost.

This model stands in contrast to other networks, such as telephony where a dropped or noisy signal cannot be readily ameliorated by packet reordering on the receiving end. Therefore the fidelity of the circuit itself, both from the perspective of throughput consistency and signal loss, is placed at greater premium.

What this demonstrates is that the choice of transport medium has considerable impact on the set of design models that a distributed system can use to solve the concurrency problem.

With \aoss there are two main dimensions: keeping physically connected but nonetheless discrete nodes concurrent, and expanding this to likewise work with nodes that are geographically dispersed. The former can be accomplished with some form of middleware that acts as reliable transport even when layered on top of a transport where no guarantee of reliable message ordering is provided.  HCP accomplishes this through the creation of a reliable messaging system layered on top of the TCP/IP backbone that connects the physical nodes that constitute the cluster. This, coupled with locking semantics among the various internal software subsystems, allows for guaranteed concurrency regardless of which node is servicing a particular read request.

Expanding this to a logical cluster dispersed across distant geographies requires the basic tradeoff of paying a premium for private network connections that come with fidelity guarantees similar to those seen in telephony, or to trade consistency for time. In the latter case, if a client requires synchronous concurrency, i.e. all nodes in the entire logical cluster that house a copy of the user data are all consistent at the same instant, the tradeoff is potentially long delays before a write success acknowledgment (ACK) can be returned to the client. Many clients aren't designed for such variable and potentially long delays in receiving a write ACK and will time out under the assumption that an error must have occurred somewhere in the system.

The alternative is to return a success ACK to the client upon the first successful write of the object and then leave it to the \ados to asynchronously ensure that a consistent view of the object is held throughout the entire object store. While this latter model introduces a level of risk and uncertainty to the client, it is generally the more economical method and can prove sufficient for applications that are unlikely to simultaneously access an object immediately after its initial ingest.

How a particular implementation of an \aos solves the two big problems of scale and concurrency has a substantial impact to the customer. Since it's comparatively easy to solve these problems at very low object counts, it can be especially challenging for the consumer of an \aos to make an informed purchasing decision as the problems that arise from these system design tradeoffs may not manifest themselves until the system has been in use for a long time and a significant object count has accumulated.

\section{DOS Characteristics} \label{s:doschars}

\subsection{Common Characteristics}
At the conceptual level all \aoss share certain similarities; they differ in the particular means of implementing these core functions, which results in differing upper bounds for performance and object count.

There's a common set of desired characteristics that a present-day enterprise
class \aos is expected to have. The minimum set includes the ability to:
\begin{itemize}
\item Grow capacity as needed;
\item Tightly couple data and metadata;
\item Present a global name space to the client;
\item Deal with a time horizon that shifts to decades; and be
\item Equally accessible over the LAN and WAN.
\end{itemize}

In the next sections we discuss these five key characteristics.

\subsubsection{Capacity on Demand}
Conceptually this ideal is simple: customers would like to purchase a system that can seamlessly grow capacity on an as-needed basis. In practice building such a system presents many challenges. The most obvious is that the hardware systems (computes and storage) will change over time. Marrying the old to the new requires at a minimum that the physical form factors are compatible. People who have owned more than one laptop in their lives know that just getting two that use the same power supply and adapter is difficult. When you multiply this problem to extend to the full complement of mechanical and electrical elements that must coexist within the same systems, it's easy to see how this problem grows geometrically.

The second, and even greater challenge, is designing a software system capable of seamless capacity additions. Object stores typically mask the detail of configuring the back-end storage subsystem from the system administrator. This proves useful in helping keep system management costs down and thus lowers the total cost of ownership. However, this also means that the \aos must continue to offer this same ability even as the underlying storage subsystems change.  Because these systems will all operate in different manners, the challenge to the \aos is to be able to know the characteristics of each back-end data store and seamlessly operate across different generations.

This causes a significant amount of processing overhead, as each device will not only have different configuration parameters but also will require different access methods to achieve optimal performance. It is up to the \aos to keep track of these details and alter the way data access is configured and realized across the spectrum of back-end stores. The result is that a single algorithm cannot be used universally for all data access methods, as it can produce inferior performance results at best and simple failure at worst.

The net result is that the variables involved with adding capacity on demand must be taken into account during the initial design of the system or the \aos will become increasingly brittle as new generations of capacity are added over time.

\subsubsection{Data and Metadata}
A significant benefit of an \aos over traditional storage is the ability to couple an arbitrary set of application- and system-defined metadata with the original data set (cf. \S\ref{s:objdef}). This allows an entirely new set of functions to be taken against objects not only at ingest, but throughout their life in the \aos as metadata presents a means for significantly expanding the value of the data.

In general the shelf life of applications will be less than that of the data stored.  Therefore, it can be critical that information that identifies the client (e.g., the revisions of the application and associated software) is stored in the metadata so that the user knows which version of the application is needed to actually \emph{use} the data when it is eventually retrieved.  Otherwise, the value of the data quickly approaches zero.  Further, metadata provides an easy mechanism for consistency since all attributes about the data can be stored in a single place readily accessible, i.e. in the metadata associated with the object.

The challenge for the \aos is to allow rich metadata that can grow and be altered post ingest, and to always be able to retain the coupling of the metadata with the object data for the entire life of the object.

\subsubsection{Global Namespace}
Another significant value of an \aos is that it presents to the clients a single global namespace. This unburdens client applications from the need to keep track of where data is stored in perpetuity, which not only simplifies the client storage logic, but also has the side effect of making applications more resilient to changes in the data center \cite{RJP2}.

\subsubsection{Time Horizon}
Traditional spinning-disk storage solutions deal with a data life measured in months or years. An \aos by contrast can be expected to deal with a data life measured in decades. The life of the objects may be dictated by regulation, or the objects may simply be expected to always be present as a matter of course.

This increase in life expectancy makes continual and automatic checks of the veracity of the stored data a must. All storage media will experience irrevocable data loss given enough time. This value, measured as mean time to data loss (MTTDL), varies by medium and usage pattern. However, the common characteristic is that for all media, MTTDL never equals zero. Therefore, it's incumbent upon the \aos to actively check and repair data objects that have become corrupted for whatever reason. Typically \aoss will use some combination of hashing and/or direct binary comparisons to guarantee that the stored data is the data actually returned to the client \cite{RJP1}.

\subsubsection{Accessibility}
A well-constructed \aos will be equally accessible by LAN or WAN. This implies that traditional LAN-based protocols such as CIFS and NFS are insufficient as the sole access mechanism. While these protocols suffice for a LAN, they're too chatty for long-distance communications. Presently, the HTTP protocol is the lingua franca of the Internet, which accounts for its prevalence in Internet-based Cloud storage systems today. However, an \aos should anticipate that the access protocol will change over time and be designed to accommodate a swap of access protocol just as it must accommodate a change in storage medium.

\subsection{Differentiation} \label{s:diffs} 

In \S\ref{s:concurrency} we mentioned that it can be difficult for an \aos consumer to know how well a system will scale or retain a consistent view across the cluster until after the system has been in use for a substantial period of time. In this section we list four areas that are readily apparent for various implementations of an object store, and therefore don't have the handicap of being seen only after the system has been running for a long time.

\subsubsection{Degree of Abstraction}
An \aos has the potential to abstract away the detail of the storage, which provides substantial benefit to both the application and the customer.  There are two main areas that can be abstracted:
\begin{itemize}
\item The ability to homogenize various back-end storage subsystems; and
\item The ability to allow arbitrary metadata to grow quite large.
\end{itemize}

The extent to which the \aos can distill away all distinction of the back-end data store has a substantial impact on the long-term usability of the system. As noted in \S\ref{s:intro}, HDD density improvements eclipse even those of processors. Therefore, a system with a time horizon of decades must provide a near-perfect level of storage abstraction, or client applications will need to change to efficiently use new storage systems and methods of data store. For example, if an \aos uses local disk within the compute nodes themselves, it must present no change to the client if the system is later swapped out to use more advanced storage subsystems.

Metadata is at the heart of any well constructed object store. The extent to which the client and system administrator are able to add unlimited custom metadata is a mark of the utility of the \aos as metadata can often grow much larger than the data itself.  Systems that limit metadata size and type invariably limit the set and type of applications that can benefit from the object store.  Additionally, it's useful to allow changes to metadata over time to extend the usefulness of the system.  For example, an fMRI scan will not change, but it's valuable to be able to update the metadata associated with the fMRI to indicate the progress of the patient, such as with a record of who the patient has seen, whether he's had surgery, etc.

\subsubsection{Discernible Namespace}
Objects stores tend to present either a completely opaque namespace, such as in the case of CAS systems that use a hash of the object as the only handle, or a namespace that is traversable and human readable, such as through the presentation of a file system facade.

The value of the former is that there can be only one handle for an object regardless of where it exists in the object store, assuming that hash collisions are not a factor. The disadvantage is that such a model makes it difficult, if not impossible, for either the client application or the system administrator to trace objects stored in the cluster.  There's a certain trust factor that comes into play when using a system that provides zero visibility into objects once they're ingested into the object store. This weakness is removed when instead the \aos presents a traversable file system facade. In this case both client application and human users can readily see the objects that are present on the cluster.

\subsubsection{Degree of Freedom}
While there are active efforts to create a standard interface for objects stores \cite{snia}, today most object stores, whether in the cloud or enterprise, create a private API that client applications must write to in order to make use of the object store. The degree to which this API is unique to one and only one vendor is the degree to which customers are locked in to that vendor's offering. While such ``stickiness'' is advantageous to the vendor, it's equally limiting to the customer. To the extent that an \aos does not require a proprietary API or allows access through multiple standard protocols such as CIFS and NFS, the customer has a greater degree of freedom to change vendors as they desire.

\subsubsection{Data Protection}
The basic tradeoff for data protection is coding complexity versus storage efficiency. The easiest system to build will simply create $n$ clones of the original object. The disadvantage of this model is that it's space inefficient; i.e., at best a customer can use only 50\% of the raw capacity.  More space-efficient means, such as RAID, have the benefit of allowing the customer to use more of the capacity purchased, but are harder to implement.

Ideally the \aos will allow customers to select the data protection model that best suits their needs and runs equally well regardless of the data protection selected. To do this well typically requires the use of a sophisticated back-end storage subsystem.

While there are means to gain greater storage efficiency without traditional RAID storage subsystems, such as erasure encoding \cite{Guru}, these methods are still relatively novel and therefore have not undergone the rigor of extensive customer use typified by RAID models.  If the \aos is to be used as the final home for object data, the means of data protection must be solid or it risks data loss.


\section{Futures} \label{s:futures} 

\subsection{Rise of Intelligent Storage}
In \S\ref{s:sysview} we introduced the notion that an \aos is in part a move from what is traditionally labeled as ``dumb storage'' to an intelligent system, capable of performing arbitrarily complex functions against the data set. This represents a significant shift in the storage industry. Previously the intelligence was left to the application and the scope of the storage subsystem was constricted to concerns such as data protection models. It's this shift to intelligent storage that has marked the method selected by new entrants to cloud storage. It's not surprising that the first to embrace intelligent storage are those outside the mainstream storage industry, as new entrants don't have existing storage lines that could be cannibalized as a result.

\subsubsection{Granularity}
State of the art today is for the scope of the intelligence to be with the \aos itself, and this is working well with object counts that number in the 100's of millions to about a billion. However, the scale issue will only grow worse.  It's a lot easier to store a petabyte than is is to store a billion objects.

To make that significant next jump in scale will likely require that the intelligence be ingrained in the objects themselves. In such a model individual objects would have the ``DNA'' to know when to create clones of themselves and how to adjust to changes in environment. For example, in the case of a rush of read requests in a particular geography, objects would be cloned and migrate to the hot spot to service requests locally. Once read activity subsided, objects would know to die off as there would no longer be a need for such a large population.

\subsubsection{Extreme Scale}
We can use other systems with very large scale as a means of comparison, such as the human organism, which contains 10's of trillions of cells. The human organism couldn't operate at such scale if it were bounded by the limit of a master control program that acted as gatekeeper to all cellular activity.  Instead, the human organism is controlled by a set of autonomic functions that operate independently of conscious thought and thus can perform the myriad functions necessary to keep such a complex of cells operating as a single unit.  It's not hard to imagine that to achieve extreme scale in the 10's of trillions, that intelligent \aoss will likewise need to push down some of the intelligence to the objects themselves, thus creating ``intelligent objects''.

In medicine similar ideas exist in the research community --- such as using nanomedicine as a novel way of targeting cancer by viewing the human organism as a system of interacting molecular networks and targeting disruptions in the system with nanoscale technologies \cite{Heath}.

Conceptually similar research is occurring in the computer science realm with protein-based computers as future replacements for existing silicon-based systems, modeling the complex protein-signaling networks that sense a cell's chemical state and respond appropriately \cite{Proteins}. Here too the idea is that to get to that next level of extreme scale means a shift away from the ideas of central processing units in hardware, or master control programs of some variety in software, to a much more granular level of actions and knowledge at the individual object level.

Perhaps the best contemporary example we see of a very large-scale system that distributes intelligence to the nodes and operates without a master control program is the Internet itself.  It's easy to understand that a system of this scale couldn't operate without distributing intelligence.

\subsection{Beyond Archive}
For traditional storage vendors there are two near-term challenges:
\begin{enumerate}
\item  Figuring out how to position object stores, and
\item Moving beyond the archive.
\end{enumerate}

While early entrants in cloud storage don't need to contend with how to position an \aos against other lines of storage, such is not the case with storage vendors themselves. It's always a challenge figuring out how to slot in new technologies in a way that is easy to explain to a global sales force while not cannibalizing sales of other lines. The method of choice thus far has been to artificially characterize \aoss as suitable only for archive data, leaving existing lines to handle other types of data. While this may solve a near-term problem of product positioning, it fails to take full advantage of the broad set of functions that an \aos is capable of performing.

This works well so long as everyone agrees to play by the same rules, which to date has been more or less true with traditional storage vendors. However, the new entrants in cloud storage aren't constraining themselves in this manner and therefore are expanding the use case for \aoss beyond just archive. The challenge for existing storage vendors is to see that the real competition in the 21st century may not come from the same competitors of the last decade --- a lesson that the now defunct minicomputer manufacturers were never able to fully learn when the microcomputer took over in the 1990s. \cite{clayton}


\section{Summary} \label{s:summary} 

In this paper we described the fundamental principles of operation of \adoss in general, with a focus on the various challenges system designers face and the associated tradeoffs inherent to design selection. We then reduced these general concepts to specific examples of the challenge of creating an \aos that is truly scalable while remaining coherent. Finally, we concluded with thoughts on how such systems must evolve to make the next big step in system scale and operation, thereby extending the market opportunity.

\section*{Acknowledgments}

I would like to thank the following reviewers: Carl D'Halluin, Jonathan Chinitz, John Dicker, John Hilliar, Scott Nyman, and Scan Putegnat.

\section*{Author}

Robert Primmer works at Hitachi Data Systems (HDS) in the Global Solutions Strategy and Development division as Sr. Technologist and Sr. Director of Content Services Product Management, where he works on the Hitachi Content Platform (\href{http://www.hds.com/products/storage-systems/content-platform/index.html}{HCP}) --- a distributed object store. Prior to HDS, he worked on the Centera and Atmos object stores at EMC.  He is a member of the ACM, IEEE, and IEEE Computer Society.


\ifpdf
\pdfbookmark[1]{References}{references}
\fi


\begin{thebibliography}{30}

\small

\bibitem{EA} A.Grunbacher, ``POSIX Access Control Lists on Linux," SuSE Labs,
  s.a.

\bibitem{RAID} A.Thomasian, M.Blaum, ``Higher Reliability Redundant Arrays:
  Organization, Operation and Coding,'' \emph{ACM Transactions on Storage}, Vol
  5., No. 3, Article 7, November 2009.

\bibitem{clayton} C.Christensen, ``The Innovator's Dilemma: The Revolutionary
  Book that Will Change the Way You Do Business,'' \emph{HarperBusiness}, 2000.

\bibitem{sciam1} C.Walter, ``Kryder's Law,'' \emph{Scientific American},
  pp. 32-33, August 2005.

\bibitem{sciam2} C.Walter, ``Letters,'' \emph{Scientific American}, p. 14,
  December 2005.

\bibitem{Telehealth} D.Josephy, ``Medicine's Next Big Battlefield: Your Home,''
  \emph{BusinessWeek}, April 27, 2009.

\bibitem{RDS} G.Chockler, et al., ``Reliable Distributed Storage,'' \emph{IEEE
    Computer}, pp. 60-67, April 2009.

\bibitem{economy} H.Sirkin, ``Slow Economy, Advancing at Warp Speed,''
  \emph{BusinessWeek}, September 8, 2009.

\bibitem{idc1} IDC, ``The Diverse and Exploding Digital Universe,'' March 2008.

\bibitem{UmbrellaFS} J.Garrisson, A.L.Narashima Reddy, ``Umbrella File System:
  Storage Management across Heterogeneous Devices,'' \emph{ACM Transactions on
    Storage}, Vol. 5, No. 1, Article 3, March 2009.

\bibitem{gray} J.Gray, ``The Transaction Concept: Virtues and Limitations,''
  \emph{Proceedings of Seventh International Conference on Very Large
    Databases}, September 1981. 

\bibitem{Heath} J.Heath, M.Davis, L.Hood, ``Nanomedicine Targets Cancer,''
  \emph{ Scientific American}, pp. 44-51, February 2009.

\bibitem{tape} J. Van Bogart, ``What can go wrong with magnetic media,''
\emph{Publishing Research Quarterly}, Vol. 12, No. 4, pp. 65-77, December 1996.

\bibitem{optical} M.Ahmetovic, et al., ``Optical Storage Media Industry
  Analysis,'' \emph{Optical Storage Media Industry Report-1}, s.a. 

\bibitem{Ratner} M.Ratner, ``Hitachi Content Archive Platform: Architecture
  Overview and Interface Performance Version 2.6, Architecture Guide and
  Performance Brief,'' June 2009.

\bibitem{Proteins} N.Ramakrishnan, U.Bhalla, J.Tyson, ``Computing with
  Proteins," \emph{IEEE Computer}, pp. 47-56, January 2009.

\bibitem{DNS} P.Vixie, ``What DNS Is Not,'' \emph{Communications of the ACM},
  Vol. 52, No. 12, pp. 53-47, December 2009.

\bibitem{REST} R.Fielding, ``Architectural Styles and the Design of
  Network-based Software Architectures,'' PhD thesis, University of California,
  Irvine, 2000.

\bibitem{RJP1} R.Primmer, C.D'Halluin, ``Collision and Preimage Resistance of
  the Centera Content Address,'' June 2005.

\bibitem{RJP2} R.Primmer, ``Efficient Long-Term Data Storage Utilizing Object
  Abstraction with Content Addressing,'' July 2003.

\bibitem{t10} R.Weber, ``Information Technology - SCSI Object-Based Storage
  Device Commands - 2 (OSD-2),'' Revision 4, July 2008.

\bibitem{snia} SNIA, ``Cloud Data Management Interface,'' Version 1.0g,
  February 9, 2010.

\bibitem{Venti} S.Quinlan, S.Dorwards, ``Venti: A new approach to archival
  storage,'' \emph{Usenix Conference on File and Storage Technologies}, 2002.

\bibitem{Rhea} S.Rhea, et al., ``Fast, Inexpensive Content-Addressed Storage in
  Foundation,'' \emph{Proceedings of the 2008 USENIX Annual Technical
    Conference}, 2008.

\bibitem{Akamai} T.Leighton, ``Improving Performance on the Internet,''
  \emph{Communicaitons of the ACM}, Vol. 51, No. 2, February 2009, pp. 45-51.

\bibitem{arr} US House of Representatives, ``Conference Report on H.R. 1,
  American Recovery and Reinvestment Act of 2009'', Feb 12, 2009, p. H1337.

\bibitem{Guru} V.Guruswami, A.Rudra, ``Error Correction up to the
  Information-Theoretic Limit," \emph{Communications of the ACM}, Vol. 52,
  No. 3, pp. 87-95, March 2009.

\bibitem{Val} V.Henson, ``The code monkey's guide to cryptographic hashes for
  content-based addressing," \emph{LinuxWorld}, November 12, 2007.

\bibitem{Cumulus} V.Vrable, S.Savage, G.Voelker, ``Cumulus: Filesystem Backup
  to the Cloud,'' \emph{ACM Transactions on Storage}, Vol 5., No. 4., Article 14,
  December 2009.

\bibitem{aws} W.Vogels, ``Eventually Consistent,'' \emph{Communications of the
    ACM}, Vol. 52, No. 1, pp. 41-44, January 2009. 

\end{thebibliography}


\end{document}